\documentclass[12pt,twoside]{article}
\usepackage{fleqn,espcrc1}

\usepackage{epsf}

\def\beq{\begin{equation}}
\def\eeq{\end{equation}}
\def\bea{\begin{eqnarray}}
\def\eea{\end{eqnarray}} 

\def\eqref#1{eq.~(\ref{eq:#1})}

\newcommand{\AmS}{{\protect\the\textfont2
  A\kern-.1667em\lower.5ex\hbox{M}\kern-.125emS}}

\title{Three-Body System with Short-Range Interactions}

\author{J. Gegelia\address{Department of Physics, Flinders University, 
Bedford Park, SA 5042 Australia}%
        \thanks{This work was carried out whilst the author was a recipient of
an Overseas Postgraduate
Research Scholarship and a Flinders University Research Scholarship
at the Flinders University of South Australia.}}
       
\begin{document}

\maketitle

\begin{abstract}
Within the framework of non-relativistic scalar effective
field theory it is shown that the problem of the cutoff dependence of the
leading order amplitude 
for a particle scattering off a two-body bound state can be solved without
introducing three-body forces. 
\end{abstract}

\medskip
\medskip

Applications of effective field theory (EFT) to problems of nuclear
physics  have been under intensive investigations during the last few years.
A review of recent developments (and references to the relevant papers) can be
found in \cite{bira}.

Generalisation of the EFT program to the three-body problem is not
straightforward. In bosonic systems and in some fermionic channels 
one
encounters a non-trivial problem. While each leading order three-body diagram
with 
re-summed two-body interactions is individually finite, the whole amplitude
shows sensitivity to the ultraviolet cutoff. 
In \cite{stoogeletter} it was argued that the addition of an one-parameter
three-body force counter-term at leading order is {\it necessary and
sufficient} to
eliminate this cut-off dependence.
 

The present paper considers the simple case of a non-relativistic
scalar particle scattering off a two-body bound state
and provides a solution of the above mentioned problem of sensitivity
to the ultraviolet cut-off without introducing three-body forces into the
leading order Lagrangian. 

\medskip


The Lagrangian of the considered EFT of non-relativistic
self-interacting boson $\phi$
is given by the following expression \cite{more3stooges}:
\begin{equation}
\label{lag}
{\cal L}  =  \phi^\dagger
            \left( i\partial_{0}+\frac{\vec{\nabla}^{2}}{2M}\right)\phi
 - \frac{C_0}{2} (\phi^\dagger \phi)^2
 - \frac{D_0}{6} (\phi^\dagger\phi)^3 + \ldots ,\nonumber
\end{equation}
where the ellipsis stands for terms with more derivatives and/or fields.
Terms with more derivatives are suppressed at low
momentum and terms with more fields do not contribute to the
three-body amplitude.
For the sake of convenience \cite{transvestite}
one can rewrite this theory
introducing a dummy field $T$ with the quantum
numbers of two bosons (referred to as ``dimeron'' \cite{more3stooges}),
\begin{eqnarray}
\label{lagt}
{\cal L}  &=&  \phi^\dagger
             \left( i\partial_{0}+\frac{\vec{\nabla}^{2}}{2M}\right)\phi
         + \Delta T^\dagger T
  -\frac{g}{\sqrt{2}} (T^\dagger \phi\phi +\mbox{h.c.})\nonumber\\
      &+&h T^\dagger T \phi^\dagger\phi
 +\ldots
\end{eqnarray}
\noindent
Observables depend on
the parameters of Eq. (\ref{lagt}) only through the combinations
$C_0\equiv  g^2/\Delta = 4\pi a_2/M$
and $D_0\equiv -3hg^2/\Delta^2$.

\begin{figure}[t]
\begin{center}
\epsfxsize=14cm\epsfbox{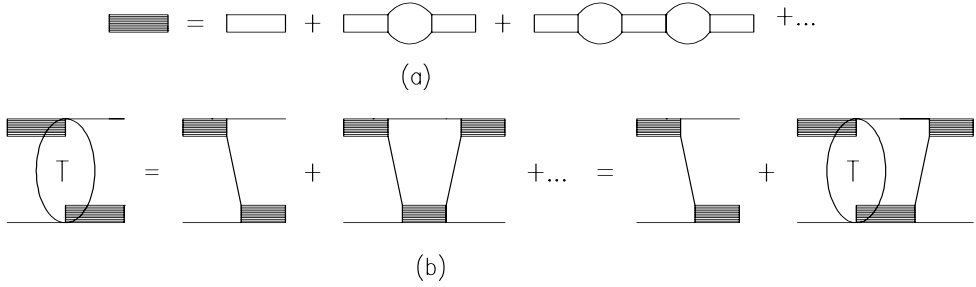}
\end{center}
\vspace*{-1cm}
\caption{{\it (a) Dressing of the dimeron. (b) Diagrams contributing to
the particle - bound-state scattering.}}
\label{fig1}
\end{figure}

The (bare) dimeron propagator is a constant $i/\Delta$ and the particle
propagator 
is given by the usual non-relativistic expression 
$i/(p^0-p^2/2M)$. The dressing of the dimeron propagator is given in
FIG.\ref{fig1} (a). Summing loop-diagrams, subtracting divergent integral at
$p^0=\vec{p}^2=0$ and removing the cut-off one gets the following dressed
dimeron propagator: 

\begin{equation}
\label{Dprop}
i S(p) =  \frac {1}{- \Delta^R
             + \frac{M g^{2}}{4\pi}
               \sqrt{-M p^0+\frac{\vec{p}^{\,2}}{4}-i\epsilon} +i\epsilon} .
\end{equation}
\noindent
Where $\Delta^R$ is the renormalised parameter ($\Delta$ has absorbed the linear
divergence). 

Standard power counting shows that diagrams which contribute to leading order
calculations of particle - two-body bound state scattering are those illustrated
in 
FIG.\ref{fig1} (b).
The sum of all these diagrams satisfies the equation
represented by the second equality in FIG.\ref{fig1} (b)
\cite{skorny},  \cite{1stoogetoo,2stooges,3stooges}:

\begin{equation}
\label{aeq}
a(p,k)
=K(p,k)+\frac{2\lambda}{\pi}\int_0^\infty dq\ K(p,q)
\frac{q^2}{q^2-k^2-i\epsilon} a(q,k),
\end{equation}
\noindent
where $k$ ($p$) is the incoming (outgoing) momentum,
$M E = 3k^2/4 - 1/a_2^2$
is the total energy, $a(p=k,k)$ is the scattering amplitude, $a_2$ is the
two-particle scattering length, and

\begin{equation}
K(p,q)= \frac{4}{3}\left(\frac{1}{a_2}+\sqrt{\frac{3}{4}p^2-M E}\right)
   \frac{1}{pq}{\rm ln}
    \left(\frac{q^2+p q+p^2-M E}
               {q^2-q p+p^2-M E}\right)
\end{equation}
Eq. (\ref{aeq}) was first derived by Skorniakov
and Ter-Martirosian (S-TM) \cite{skorny} and has $\lambda=1$ for the boson case.
Three nucleons in the spin $J=1/2$ channel obey a pair
of integral equations with similar properties to this bosonic equation.

It was shown in \cite{danilov} that for $\lambda =1$ the homogeneous equation
corresponding to Eq. (\ref{aeq}) has a solution for arbitrary $E$. This solution
is 
well-defined except for a normalisation constant and hence the solution of
Eq. (\ref{aeq}) contains an arbitrary parameter. The sum of
the diagrams in FIG.\ref{fig1} (b) is only one of the solutions. 
Hence, given the general solution of Eq. (\ref{aeq}), to find this sum 
one would have to fix the value of the arbitrary parameter appropriately.  

The fact that the homogeneous equation corresponding to Eq. (\ref{aeq}) has a
solution for arbitrary $E$ is not surprising: since Eq. (\ref{aeq}) corresponds
to a coordinate space $\delta $-function potential, the use of the Thomas
theorem \cite{thomas} 
combined with the Efimov effect \cite{efimov} explains the existence of
solutions for arbitrary $E$.
Note that two-body forces are not actually of zero range in EFT. Although
Eq. (\ref{aeq}) can be derived from the leading order Lagrangian of EFT, this
equation is not a leading order approximation of a more general equation: 
there are no
consistent equations for renormalised amplitudes in EFT if the cut-off is
removed after renormalization. The problem is that EFT is a
non-renormalizable theory in the traditional sense and hence to remove all
divergences which occur in the equations for amplitudes one would need to
include contributions of an infinite number of counter-terms at any finite order
(except perhaps leading order)
approximation. 
Hence EFT with removed cut-off describes the 
amplitude for a particle scattering off a two-body bound state as a sum of an
infinite number of diagrams. The EFT approach is concerned with Eq. (\ref{aeq})
only because one of its solutions corresponds to this sum of diagrams.


A great advantage of cut-off theory is that one can write down consistent
equations, and the solutions of these equations are equivalent to the
renormalised (with removed cut-off) amplitudes up to the order one is working
with. If working with equations of cut-off theory it is necessary to keep
the cut-off finite even though at leading order the cut-off can be removed,
giving Eq. (\ref{aeq}). As the 
equations with finite cut-off do not correspond to any system with local
($\delta $-function type) potential, there are no three-body bound states with
arbitrarily large negative energies. The solution of the homogeneous equation
corresponding to equation (\ref{aeq}), which exists for any value of the
energy, does not carry any physical information. The existence of this solution
is a result of the
incorrect procedure of removing the cut-off in the leading order
equations of the cut-off
theory. 
Note that the amplitude
determined  from the equation of cut-off theory can contain some
non-perturbative contributions in addition to the sum of the infinite number of
diagrams drawn in FIG.1 (b) but these
non-perturbative effects can not have anything to do with non-physical solutions
of the homogeneous equation.    

One can still use Eq. (\ref{aeq}) to find the amplitude for a particle 
scattering off two-body bound state, but one should keep in mind that it
contains non-physical information encoded in the solution of the corresponding
homogeneous equation. 

As will be seen below the EFT approach fixes uniquely the arbitrary parameter
present in the general solution of Eq. (\ref{aeq}). This particular
solution with an
appropriately fixed value of the arbitrary parameter is the
scattering amplitude.

One can study the asymptotic behaviour of $a(p,k)$ for large $p$.
Up to terms decreasing as $p^{-1}$ the function $a(p,k)$ has the form
\cite{danilov}:
\begin{equation}
a(p,k)\sim \sum_i A_i\left( k\right)p^{s_i}
\label{aasymptotic}
\end{equation}
where $s_i$ are roots of the following equation:

\begin{equation}
\label{seq}
1- \frac{8\lambda}{\sqrt{3}}
   \frac{\sin\frac{\pi s}{6}}{s \cos\frac{\pi s}{2}}=0.
\end{equation}
\noindent
The summation in Eq. (\ref{aasymptotic}) goes over all solutions of
Eq. (\ref{seq}) for which $|{\rm Re} s|<1$.
Eq. (\ref{seq}) has two roots for which $|{\rm Re} s|<1$:
$
s=\pm is_0, \ \ s_0\approx 1
$.
Hence, Eq.(\ref{aasymptotic}) becomes:
\begin{equation}
a(p,k)\sim A_1\left( k\right)p^{is_0}+A_2\left( k\right)p^{-is_0}
\label{aasymptotics}
\end{equation}
One of the arbitrary constants $A_1\left( k\right)$ and $A_2\left( k\right)$
is determined by the other when this solution is joined to the solution in the
region of small $p$. Hence the solution of Eq. (\ref{aeq}) depends on a single
arbitrary parameter. The asymptotic behaviour of the solution of the homogeneous
equation 
corresponding to Eq. (\ref{aeq}) is evidently the same.

Iterating the equation (\ref{aeq}) one gets a series which is
equivalent to the sum of the diagrams in FIG.\ref{fig1} (b). As $s_0$ does not
have an expansion
in $\lambda$, it should be clear that for the
sum of the considered diagrams (if it exists) the parameters
$A_1\left( k\right)$ and $A_2\left( k\right)$ must be vanishing. 


Hence the EFT with removed cut-off supports the conclusion drawn from general
considerations, namely
that the non-physical solution of the homogeneous equation has to be
eliminated.

To find the sum of the considered infinite number of diagrams one needs to
construct  
a solution with non-oscillating asymptotic behaviour i.e. with vanishing 
$A_1(k)$ 
and $A_2(k)$. Note that there is only one solution with such asymptotic
behaviour.   


To summarise, the leading order EFT 
for a spinless
particle scattering off a two-body bound state leads to the
equation 
of S-TM together with a boundary condition at the
origin 
(in configuration space) which eliminates the oscillating behaviour. Hence EFT
resolves quite naturally
the problem of the choice for the arbitrary parameter present
in the general solution of this equation.

\end{document}